\documentclass[letterpaper]{article}
\usepackage{fullpage}
\usepackage{graphicx}
\usepackage{amssymb}

\begin{document}

\null
\vskip-2\baselineskip
\begin{flushright}
UB-HET-08-02\\
July 2008
\end{flushright}

\bigskip

\begin{center}
\Large
Perturbative QCD at High Energy Colliders\footnote{
Contribution to the Symposium: 50+ Years of High Energy Physics at UB,
University at Buffalo, October 20-21, 2006. Published in Int.\ J.\
Mod.\ Phys.\ E {\bf17} (2008) 870-890.}

\bigskip
\bigskip

\large
Richard J. Gonsalves\footnote{Email: phygons@buffalo.edu}\\
\normalsize\sl
Department of Physics\\
University at Buffalo, The State University of New York\\
Buffalo, NY 14260-1500, USA
\end{center}

\bigskip

\begin{abstract}
Selected applications of perturbative Quantumchromodynamics (QCD) to
predictions of the Standard Model for processes at high energy
colliders are reviewed with emphasis on past successes and future
problems.  This is a personal retrospective is not intended to be a
comprehensive review of the field.
\end{abstract}

\section{Introduction}

It is with great pleasure that I dedicate this contribution to
Professor Piyare Lal Jain, a pioneering faculty member in experimental
high energy and heavy ion physics at the University at Buffalo over
the past fifty years, and a respected senior colleague of mine in the
Department of Physics over the past twenty-five years.  His mentoring,
friendship and collaboration\cite{Jain:1997pc} have been invaluable to
me.

\section{Renormalization Prescription Dependence of QCD Predictions}

My earliest work in the field of perturbative QCD with
Celmaster\cite{Celmaster:1979dm,Celmaster:1979km} was on the
renormalization prescription of the effective coupling $\alpha_s$ and
its experimental consequences.

Quantumchromodynamics (QCD) is an asymptotically-free quantum field
theory.  The scaling behavior of Green's functions in the deep
Euclidean region can be computed as a perturbation series in the
renormalized coupling constant $\alpha_s$, which depends on the
characteristic energy scale involved in the process.  This property of
QCD was first used to obtain predictions for scale breaking of the
naive parton model in deeply-inelastic lepton-hadron scattering.  The
domain of applicability of perturbative QCD has since been extended to
include $e^+e^-$ annihilation to hadrons and numerous other infrared
safe observables at high energy colliders\cite{EllisStirlingWebber}.

A typical prediction of perturbative QCD is an $n$-th moment of a
structure function in deeply-inelastic scattering\cite{Bardeen:1978yd}
\begin{equation}
M_n=g^{a_n}[1+b_ng^2+{\cal O}(g^4)]A_n\;,
\label{Mng-eq}
\end{equation}
where $g=\sqrt{4\pi\alpha_s}$ is the SU(3)$_c$ coupling parameter,
$a_n$ and $b_n$ are calculable numbers, and the normalization $A_n$
cannot be calculated in perturbation theory.
The coefficient $b_n$ depends on how ultraviolet divergences are
renormalized, and so implicitly does the coupling $g$.
The renormalization-prescription dependence of $b_n$ must be related
to that of $g$ because physical observables cannot depend on how the
theory is renormalized.
To exhibit this dependence, express $g$ in terms of the coupling $g'$
in a different renormalization scheme:
\begin{equation}
g=g'[1+ag'^2+{\cal O}(g'^4)]\;,
\label{ggprime-eq}
\end{equation}
where $a$ is a calculable constant.
To obtain the coefficient $b_n'$ in the second scheme, substitute
the expression for $g$ in Eq.~(\ref{ggprime-eq}) in Eq.~(\ref{Mng-eq}):
\begin{equation}
M_n=(g')^{a_n}[1+(b_n+aa_n)g'^2+{\cal O}(g'^4)]A_n
=(g')^{a_n}[1+b'_ng'^2+{\cal O}(g'^4)]A_n
\;.
\label{Mngprime-eq}
\end{equation}

If the perturbation series in Eqs.~(\ref{Mng-eq},\ref{Mngprime-eq}) were
known to all orders in the couplings $g,g'$, then the
renormalization-prescription dependence would have no observable
consequences.
In practice, the series cannot be computed beyond the first two or
three orders of perturbation theory for most observables.
The truncated series do depend on renormalization prescription, and
differ by terms of the leading uncomputed order in the series.
Since $\alpha_s\sim0.1{-}0.3$ at currently-accessible energy scales,
the discrepancies are experimentally significant.

This type of dependence on the re-definition of a small expansion
parameter will arise in any truncated series expansion.  In
particular, perturbative series in the fine structure constant
$\alpha$ of Quantumelectrodynamics (QED), are also subject to
truncation ambiguities.  The issue is less severe in QED however for
two reasons: First, the expansion parameter $\alpha\simeq1/137.034$ is
much smaller than $\alpha_s$, and so are truncation ambiguities.
Secondly, there is a naturally preferred definition of $\alpha$ which
is experimentally measurable in principle.  In fact, the cross section
for Compton scattering of a photon from a spin one-half charged
particle can be shown to reduce to the classical Thompson cross
section to all orders in perturbation theory in the limit of zero
photon energy.  There is no such natural definition in QCD, either in
the low-energy limit where the theory becomes strongly interacting, or
in the high-energy limit where the effective coupling tends to zero
due to asymptotic freedom.

Several approaches have been proposed to deal with the prescription
dependence of predictions in perturbative QCD:

\begin{itemize}

\item {\bf Good universal prescriptions:}
The strategy is to find and use a prescription that appears to give
a reasonably convergent perturbation series in a wide variety of
processes.
The archetypical example of this approach is the widely-used
$\overline{\rm MS}$ scheme recommended by Bardeen, Buras, Duke and
Muta\cite{Bardeen:1978yd} in which
dimensionally-regularized
quantities are renormalized by subtracting poles and certain
associated constants in the combination
\begin{equation}
\frac{2}{4-N}+[\log(4\pi)-\gamma_{\rm E}] \;,
\end{equation}
where $N$ is a continuation of 4 space-time dimensions and $\gamma_{\rm E}$
is Euler's constant.
This is a modification of the minimal subtraction (MS)
scheme of t'Hooft and Veltman\cite{'t Hooft:1972fi} in which
only the pole in $N-4$ is subtracted.
Another proposal by Brodsky, Lepage and Mackenzie\cite{Brodsky:1982gc}
uses light-quark vacuum polarization
insertions to determine the scale of $\alpha_s$ in any process that does
not involve gluon-gluon couplings in leading order.

\item {\bf Fastest apparent convergence:}
It is by no means guaranteed that a good universal prescription must
exist.
A different strategy suggested by Grunberg\cite{Grunberg:1980ja}
tailors the prescription to each
individual process in order to make the higher order terms in the
series as small as possible.
For example, in Eq.~(\ref{Mngprime-eq}), one might choose $b'_n=0$,
so that the QCD prediction is given by the leading order term.
The different coupling parameters, determined experimentally for
example from the different moments $M_n$, are related to one another
by Eqs.~(\ref{ggprime-eq},\ref{Mngprime-eq}).

\item {\bf The principle of minimum sensitivity:}
Stevenson\cite{Stevenson:1981vj} suggested choosing a renormalization
prescription optimally for
each process considered by requiring that the truncated perturbative
expression be an extremum with respect to variations in the parameters
of the renormaliztion group of Gell-Mann and Low\cite{GellMann:1954fq},
and Stueckelberg and Petermann\cite{Stueckelberg:1953dz}.
At next-to-leading order in QCD, predictions depend on one renormalization
group parameter, the scale $\mu$ of the effective coupling $\alpha_s(\mu)$.

\end{itemize}

Each of these strategies has its advantages and shortcomings.
There is no rigorous justification for any of them within the
framework of quantum field theory.
Each of them is a more or less educated guess concerning the size of
uncomputed terms in the perturbation series.
The only way to verify this guess is to compute these unknown terms!

\subsection{The momentum-space subtraction prescription}

Celmaster and Gonsalves\cite{Celmaster:1979dm,Celmaster:1979km}
proposed an alternative to the MS and $\overline{\rm MS}$
prescriptions called momentum-space subtraction (MOM).
In this approach, one considers the basic off-shell Green's functions
that must be renormalized in any perturbative calculation beyond tree
level.
At one-loop level, these are the three-gluon, the quark-gluon, and
(in covariant gauges) the ghost-gluon scattering amplitudes.
The theory is renormalized by subtracting these fundamental building-block
amplitudes at an energy scale that is typical of the process
under consideration, which also serves as the scale of the effective
coupling $\alpha_s$.

The motivation for the MOM prescription is the intuition that ultraviolet
behavior is dominated by the off-shell Green's functions that need to
be renormalized.
If these Green's functions are primarily responsible for the divergent
behavior of the unrenormalized theory, then subtracting their
contributions at the typical energy scale of the process will minimize
their effect on the perturbative correction terms.
Because subtractions are performed on a small number of generic Green's
functions, momentum-space subtraction is an example of a good universal
prescription.

The MOM prescription proposal suffers from some ambiguities.  There are
three different vertices at one-loop level that can be chosen to define
the subtraction.  Each of these off-shell vertices is gauge dependent,
and so a particular gauge must be chosen to define the subtraction.
And finally, each three-point vertex is has two independently-variable
momenta flowing through it, and these must be fixed in relation to the
energy scale.

A detailed study of these vertices showed that it is possible to define
a class of MOM prescriptions that gives reasonably-sized next-to-leading
order QCD corrections in a number of processes.
The corrections are only weakly gauge-dependent if a covariant gauge is
used with the gauge parameter allowed to vary quite freely in the range
of the usual Feynman and Landau gauges.
We choose a natural and intuitively reasonable choice of momenta at
which to subtract a three-point function, namely the symmetric
Euclidean point $p_1^2=p_2^2=p_3^2=-\mu^2$.
The off-shell Green's functions are related by Ward-Takahashi identities,
which relate the different prescription in the MOM class to one another.
For symmetric-point subtraction, the class of MOM prescriptions is also
only weakly dependent on the chosen vertex.

\subsection{The Celmaster-Gonsalves relation for $\Lambda$s in different
schemes}

Celmaster and Gonsalves\cite{Celmaster:1979km} showed that
$\Lambda$'s defined in different schemes can be related exactly by
performing a one-loop calculation.

It is well known that the renormalization group equation
\begin{equation}
\mu\frac{{\rm d}g(\mu)}{{\rm d}\mu}=\beta(g(\mu))=
-\beta_0g^3-\beta_1g^5+\ldots\;,
\end{equation}
where $\beta$ is the QCD beta function with leading
coefficient $\beta_0=11-2n_{\rm f}/3$, can be integrated to express
the scale dependence of the QCD coupling
\begin{equation}
g^2(\mu) =\frac{1}{\beta_0\log(\mu^2/\Lambda^2)}
-\frac{\beta_1\log\log(\mu^2/\Lambda^2)}{\beta_0^3\log^2(\mu^2/\Lambda^2)}
+ \ldots
\end{equation}
in terms of a renormalization-group invariant QCD parameter $\Lambda$, which
does however depend on the renormalization prescription used to define
$g(\mu)$.
This relation can be solved perturbatively for the ratio
\begin{equation}
\frac{\Lambda}{\Lambda'} = \exp\left[\frac{1}{2\beta_0}\left(
\frac{1}{g'^2(\mu)} - \frac{1}{g^2(\mu)}\right) + {\cal O}(g^2)\right] \;,
\end{equation}
where the coupling parameters are related by Eq.~(\ref{ggprime-eq}).
Because $\Lambda$ and $\Lambda'$ are scale-invariant constants, the
limit $\mu\rightarrow\infty$ can be taken on the right hand side to give
\begin{equation}
\frac{\Lambda}{\Lambda'} = \exp\left[\frac{a}{\beta_0}\right]\;,
\end{equation}
where the constant $a$ relates the two prescriptions in one-loop order.
This result is {\em exact} to all orders in perturbation theory because
of asymptotic freedom\cite{Gross:1973id,Politzer:1973fx} $g(\infty) = 0$.

The Celmaster-Gonsalves relation has been used to relate $\Lambda_{\rm
Lattice}$ as measured in lattice QCD to $\Lambda_{\overline{\rm MS}}$
determined from high energy
phenomenology\cite{Hasenfratz:1980kn,Dashen:1980vm}.

\section{QCD Radiative Corrections in Electron-Positron Annihilation}

\begin{figure}[th]
\centering{\includegraphics[width=12cm,clip]{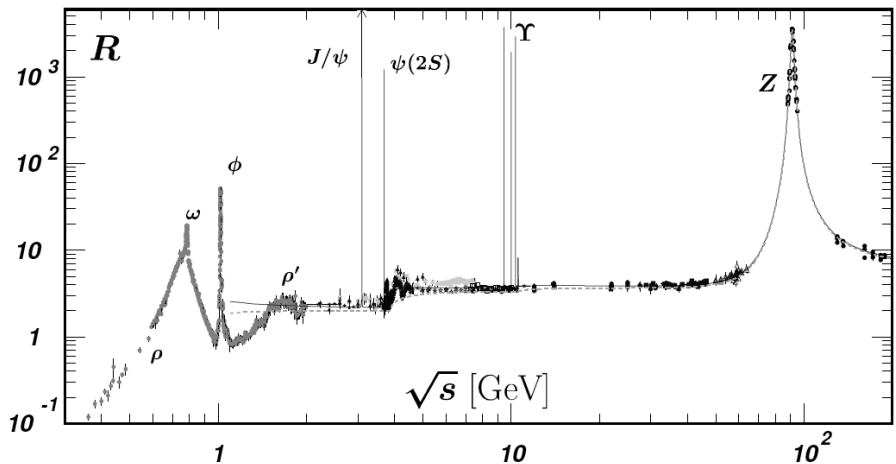}}
\vspace*{8pt}
\caption{Data
on the $R$ ratio in $e^+e^-$ annihilation from the PDG Review of
Particle Physics\protect\cite{Yao:2006px}.
}
\label{pdg-repem}
\end{figure}

On of the most powerful probes of the short-distance structure of hadrons
is the virtual photon produced by annihilation of an energetic pair in
colliding beams of electrons and positrons.
This process provides a plethora of measurable inclusive cross sections
and distributions that have been used to discover new quark species,
and to study bound states of heavy quarks, the densities of partons in
hadrons, and the fundamental structure of the
quark-gluon interaction.
The simplest observable is the total cross section for annihilation to
hadrons, which can be expressed in the form
\begin{equation}
\sigma(e^+e^-\rightarrow\hbox{hadrons}) = \frac{4\pi\alpha}{3E_{\rm cm}^2}
\left(\sum_f Q_f^2\right)\left[1+\frac{\alpha_s}{\pi} + K
\left(\frac{\alpha_s}{\pi}\right)^2 + \cdots \right] \;,
\label{sigmatoteq}
\end{equation}
where the sum is over quark flavors $f=u,d,s,c,b,t$, with
$Q_f$ being the fractional quark charge, and the next-to-leading
order coefficient $K$ is calculable in perturbative QCD.
This simple form holds in regions far from flavor thresholds with the
flavor sum being restricted to flavors with $4m_f^2<E_{\rm cm}^2$.
The cross section is proportional to the number of quark colors and
thus provides a direct measurement of the $3$ in SU(3).
The fractional electric charge of each new quark species is directly
measured by the increase of the flavor sum as the threshold is crossed,
and is most dramatically illustrated by the data
in Fig.~(\ref{pdg-repem}) on the ratio
\begin{equation}
R =\frac{\sigma(e^+e^-\rightarrow\hbox{hadrons})}{\sigma(e^+e^-\rightarrow
\mu^+\mu^-)} = 3
\left(\sum_f Q_f^2\right)\left[1+\frac{\alpha_s}{\pi} + K
\left(\frac{\alpha_s}{\pi}\right)^2 + \cdots \right] \;.
\end{equation}

The leading $\alpha_s/\pi$ correction was first computed by
Schwinger\cite{Schwinger:1948yk}.
The next-to-leading order coefficient $K$ was computed
analytically by Celmaster and
Gonsalves\cite{Celmaster:1979xr,Celmaster:1980ji},
\begin{equation}
K_{\overline{\hbox{MS}}} = \frac{365}{24} - \frac{11}{12}n_f
-2\beta_0\zeta(3) \simeq 1.986 - 0.115n_f \;,
\end{equation}
where $\zeta$ is Riemann's zeta function.
This result agreed precisely with an earlier numerical
calculation by Dine and Sapirstein\cite{Dine:1979qh}
on gauge-invariant subsets of diagrams and their sum.
The analytic result was also obtained in an independent calculation by
Chetyrkin, Kataev and Tkachov\cite{Chetyrkin:1979bj}.

The next-to-leading order coefficient can be expressed in other
renormalizaton schemes using Eq.~(\ref{ggprime-eq}):
\begin{eqnarray}
K_{\hbox{MOM}} &=& \frac{365}{24} - \frac{85}{36\sqrt3}
{\rm Cl}_2\left(\frac{\pi}{2}\right)-\frac{23}{36}n_f -
2\beta_0\zeta(3)
\simeq -2.193 + 0.162n_f \;, \\
K_{\hbox{MS}} &=& \frac{365}{24} - \frac{11}{12}n_f +
\frac{\beta_0}{2}\left[\log(4\pi)-\gamma_{\rm E}-4\beta_0\zeta(3)\right]
\simeq 7.359 - 0.441n_f \;,
\end{eqnarray}
where ${\rm Cl}$ is Clausen's function, and the MOM scheme is defined
by the quark-quark-gluon vertex in Landau gauge.
Note that MOM and $\overline{\rm MS}$ give a well-behaved perturbation
series.
The simple MS scheme is less well behaved.

The theoretical prediction for $R$ has since been extended to ${\cal
O}(\alpha_s^3)$ by Gorishnii, Kataev and Larin\cite{Gorishnii:1988bc},
and Surguladze and Samuel\cite{Surguladze:1990tg}, and work is in
progress\cite{Baikov:2006nb} on the ${\cal O}(\alpha_s^4)$
coefficient.

\subsection{The photon propagator in QCD}

The total annihilation cross section can most easily be obtained from
the 2-point correlation function for the electromagnetic vector potential
\begin{equation}
D_{\mu\nu}(q) = \int {\rm d}xe^{iq\cdot x}\langle T[A_\mu(x)A_\nu(0)]\rangle
=\frac{-i}{q^2}\left(g_{\mu\nu}-\frac{q_\mu q_\nu}{q^2}\right)D(-q^2) +
\hbox{gauge terms}\;.
\end{equation}
The total cross section is given by the imaginary part of the photon
propagator
\begin{equation}
\sigma(e^+e^-\rightarrow\hbox{hadrons}) = -\frac{4\pi\alpha}{q^2}{\Im\rm m}
D(-q^2)\;,
\end{equation}
with $q^2=E_{\rm c.m.}^2$.

If QCD with massless quarks were a finite theory, then the dimensionless
function $D(-q^2)$ would be a pure number independent of $q^2$.
Because of ultraviolet and infrared divergences,
the theory must be regularized to
compute $D(-q^2)$ as a perturbative series in the QCD coupling $\alpha_s$.
This is most conveniently done by analytically continuing $D(-q^2)$
to the region of
spacelike $q^2<0$ and to space-time dimension $N=4-\epsilon$.
The unrenormalized function $D$ is then real and finite,
with divergences manifest as poles in $\epsilon$
\begin{equation}
D(-q^2) = 1+\frac{\alpha_0}{\pi}\sum_{n=1}^\infty(-q^2)^{-n\epsilon}
d_n(\epsilon)\left(\frac{\alpha_{s0}}{\pi}\right)^{n-1}
+{\cal O}(\alpha_0^2)\;,
\end{equation}
where $\alpha_0$ and $\alpha_{s0}$ are the unrenormalized (bare or zero order)
QED and QCD coupling parameters respectively, and the coefficient
\begin{equation}
d_n = \frac{d_{n,n}}{\epsilon^n}+\frac{d_{n,n{-}1}}{\epsilon^{n{-}1}}+
\cdots+\frac{d_{n,1}}{\epsilon}+d_{n,0} + {\cal O}(\epsilon)\;,
\label{residueseq}
\end{equation}
is got by computing $n$-loop Feynman diagram contributions to the
photon propagator.  The dependence of the unrenormalized propagator
on the sole dimensional quantity $q^2$ in massless QCD
follows from dimensional analysis.

The propagator is made finite in the limit $\epsilon\rightarrow0$
by subtracting divergences at a renormalization scale $\mu$.
The bare QCD coupling $\alpha_{s0}$ is then replaced by the
effective couple $\alpha_s(\mu)$.
Since the electromagnetic current is conserved, the bare QED
coupling $\alpha_0$ can simply be replaced by its renormalized
value (which differs from the usual fine structure constant
$\alpha$ by terms of order $\alpha^2$).
It is then straightforward to express the coefficient $K$ in
Eq.~(\ref{sigmatoteq}) in terms of the residues of the poles in
Eq.~(\ref{residueseq}).

\begin{figure}[th]
\centering{\includegraphics[width=8cm,clip]{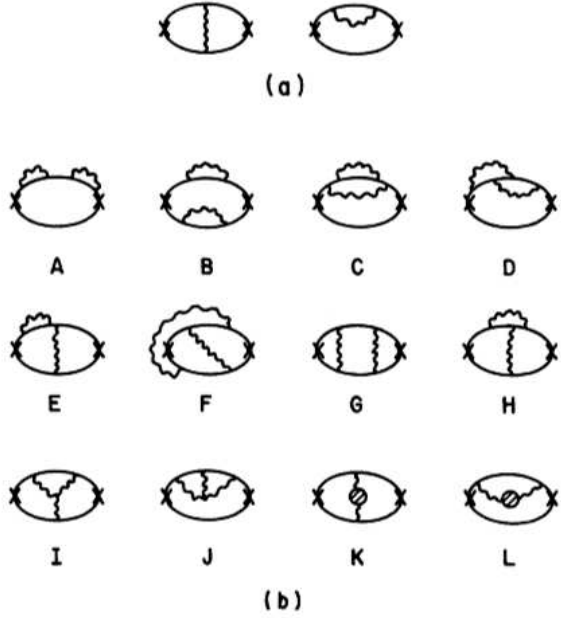}}
\vspace*{8pt}
\caption{
Feynman diagrams which contribute to the photon propagator in QCD
through order $\alpha_s^2$.
}
\label{sigmatotdiags}
\end{figure}

\subsection{The Gegenbauer expansion technique}

The residues $d_{n,i}$ in Eq.~(\ref{residueseq}) were computed by
evaluating the Feynman diagrams in Fig.~(\ref{sigmatotdiags}) assuming
massless quarks and using dimensional regularization.  The calculation
was done by generalizing a well-known Chebyshev polynomial expansion
technique developed by Rosner and
others\cite{DeRafael:1974iv,Levine:1974xh} for QED to $N$-dimensions.
The appropriate generalization is provided by Gegenbauer polynomials
$C_n^\lambda(x)$ with $\lambda=(N-2)/2$ and $x$ the cosine of the
polar angle in $N$-dimensional spherical coordinates.

The Gegenbauer polynomials\cite{Abramowitz} can be generated
from $C_0^\lambda=1$ using the recursion formula
\begin{equation}
2(n+\lambda)xC_n^\lambda(x) = (n+1)C_{n+1}^\lambda(x)
+(1-\delta_{n,0})(n+2\lambda-1)C_{n-1}^\lambda(x) \;,
\end{equation}
and obey the orthogonality relation
\begin{equation}
\int_{-1}^{+1}{\rm d}x(1-x^2)^\lambda C_n^\lambda(x) C_m^\lambda(x)
=\delta_{n,m}\frac{\pi2^{1-2\lambda}\Gamma(n+2\lambda)}{n!(n+\lambda)
[\Gamma(\lambda)]^2} \;.
\label{g-orthog}
\end{equation}

The basic idea is to expand propagator denominators of massless
quarks or gluons using an $N$-dimensional generalization of the Legendre
polynomial expansion
\begin{equation}
\frac{1}{(\vec k_1-\vec k_2)^2} = \frac{1}{k_>^2}\sum_{n=0}^\infty
f_n\left(\frac{k_<}{k_>}\right)C_n^\lambda(\hat k_1\cdot \hat k_2) \;,
\end{equation}
where the coefficient functions $f_n$ are related to hypergeometric
functions and can be expressed as a power series in the dimensional
regularization parameter $\epsilon$
\begin{equation}
f_n(x) = x^n\left[
1+\frac{\epsilon}{2}\left\{
(1-\delta_{n,0})\sum_{j=1}^n\frac{1}{j} + (n+1)\sum_{j=1}^\infty
\frac{x^j}{j(n+j+1)}
\right\} + {\cal O}(\epsilon^2) \right] \;.
\end{equation}
Integrations over loop momenta can then be performed using the
orthogonality relation in Eq.~(\ref{g-orthog}) and various other properties
of the Gegenbauer polynomials.

The calculation of Chetrykin, Kataev and Tkachov\cite{Chetyrkin:1979bj}
used a similar polynomial expansion technique in position
space\cite{Vladimirov:1979zm}.

\section{The Quark Electromagnetic Form Factor in QCD}

\begin{figure}[th]
\centering{\includegraphics[width=8cm,clip]{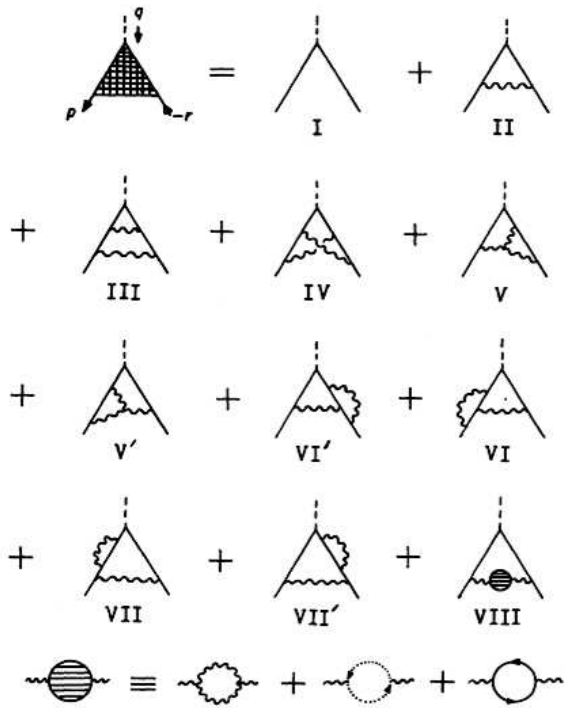}}
\vspace*{8pt}
\caption{
One-particle irreducible Feynman diagrams which contribute to the
quark electromagnetic form factor in QCD.  Solid, wavy, and dashed
lines represent quark, gluon, and ghost propagators, respectively.
Diagram IV is the only non-planar diagram in this set.
}
\label{qff-diagrams}
\end{figure}

A different method for computing the total annihilation cross section
discussed in the preceding section would be to calculate the
amplitudes for a virtual photon to decay to to 2, 3, or 4 partons,
and then sum the individual cross sections integrated over the phase
space of final state partons.  Fig.~(\ref{qff-diagrams}) shows the
Feynman diagrams which contribute to the 2-parton final states through
order $\alpha_s^2$, i.e., to the quark electromagnetic form factor.

The form factor is ultra-violet finite (when all QCD counterterms are
included) because the electromagnetic current is conserved, but it is
infrared divergent due to soft and collinear singularities.  These
singularities are canceled by contributions from 3- and 4-parton final
states so the total cross section is finite.  The form factor is
therefore an unphysical part of a measurable physical observable.
However, it is an object of great theoretical interest because the
soft and collinear singularities are generic in that they determine
double-logarithmic Sudakov effects in many measurable cross
section---see the review article by Collins\cite{Collins:1989bt}.  It
was therefore of interest to study these singularities.

A calculation\cite{Gonsalves:1983nq} of all two-loop integrals
required for an analytical calculation of the form factor was
completed in 1983, including all terms of ${\cal O}(\epsilon^{-n})$
with $0\le n\le4$.  The most challenging sets of integrals to
compute were those which contribute to the box-plus-triangle diagram
III and the non-planar diagram IV in Fig.~(\ref{qff-diagrams}).  The
calculations were done by converting the two-loop momentum integrals
to 5-dimensional integrals over Feynman parameters.  The Feynman
parameter integration were performed analytically in succession using
various techniques including infinite-series expansion of binomial
expressions where necessary.  Care had to be taken to ensure that the
binomial expansions converged after each successive integration: where
necessary, the integration interval was subdivided and appropriately
converging expansions used in each region.

The results for the two-loop scalar integrals were verified by van
Neerven\cite{van Neerven:1985xr} and by Kramer and
Lampe\cite{Kramer:1986sr}.  Gehrmann, Huber and
Maitre\cite{Gehrmann:2005pd} have recently obtained elegant exact
expressions for the integrals in diagram IV in $N$ dimensions.

These two-loop results have been used in several ${\cal
O}(\alpha_s^2)$ QCD calculations including the Drell-Yan total cross
section by Hamberg, van Neerven and Matsuura\cite{Hamberg:1990np}, the
total cross section for Higgs production by Harlander and
Kilgore\cite{Harlander:2002wh} and Anastasiou and
Melnikov\cite{Anastasiou:2002yz}, the Drell-Yan rapidity distribution
by Anastasiou, Dixon, Melnikov and Petriello\cite{Anastasiou:2003ds},
and the Higgs total cross section and Drell-Yan and Higgs rapidity
distributions by Ravindran, Smith and van
Neerven\cite{Ravindran:2003um}.

In a closely related calculation\cite{Gonsalves:1986af}, a tensor
decomposition scheme that is exact in $N$ dimensions was used to
compute the one-loop QCD corrections to the $e^+e^-$ total
annihilation cross section.  Unlike the process $\gamma^*\rightarrow
q\bar q$ in which there is a single scalar form factor, the process
$\gamma^*\rightarrow q\bar qg$ involves 6 independent invariant
tensors and corresponding form factors in $N=4$ dimensions.  When the
amplitude is continued to $N$ dimensions, there is a set of 17
linearly-independent tensors and corresponding form factors into which
it can conveniently be decomposed.  The decomposition makes it
possible to compute the dimensionally-regularized amplitude very
efficiently in ${\cal O}(\alpha_s)$, and the technique readily extends
to two or more loops.

Over the past few years, a number of results on form factors have
appeared in the literature, including analytic results for arbitrary
masses and momenta in QED\cite{Mastrolia:2003yz,Bonciani:2003ai} and
in QCD\cite{Bernreuther:2005gq,Bernreuther:2005rw}, and impressive new
results on massless quark and gluon form factors at three
loops\cite{Moch:2005tm}.

\section{Chromo-Electroweak Interference and Parity Violation}

\begin{figure}[th]
\centering{\includegraphics[width=8cm,clip]{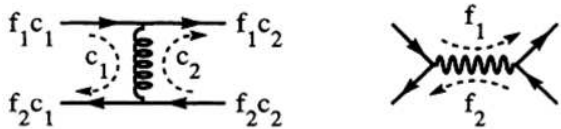}}
\vspace*{8pt}
\caption{
A solid line represents a quark or
antiquark with color $c$ and flavor $f$.  The wavy line represents a
color-neutral electroweak boson and the curly line a flavor-neutral
gluon.  The two amplitudes can interfere because their initial states
(on the left) can have identical quantum numbers as can their final
states (on the right).
}
\label{cewi-int}
\end{figure}

\begin{figure}[th]
\centering{\includegraphics[width=8cm,clip]{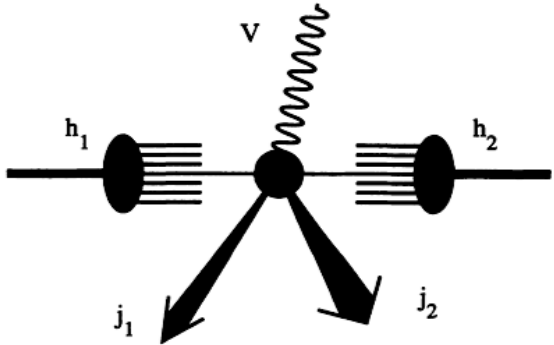}}
\vspace*{8pt}
\caption{
Hadrons $h_1$ and $h_2$ collide to produce an on-shell electroweak
boson $V$ and two hadron jets $j_1$ and $j_2$ at large transverse
momentum.
}
\label{cewi-had}
\end{figure}

In the Standard Model the process $q\bar{q}\rightarrow q\bar{q}V$
(where $V=W^\pm,Z^0$ or $\gamma$) can occur via gluon exchange and
also via $W^\pm$ or $Z^0$ exchange.  The corresponding chromodynamic
and electroweak amplitudes can interfere with one
another\cite{Gonsalves:1985jn} as shown in Fig.~(\ref{cewi-int}).
These interference cross sections are largest when the exchanged
$W^\pm$ or $Z^0$ is on-shell when they are also odd under parity.
Interference cross sections computed using helicity-amplitude
techniques were analyzed\cite{Gonsalves:1993sz} for all interesting
subprocesses as well as for the processes $q\bar{q}\rightarrow
q\bar{q} l\bar{l}$ in which the lepton pair $l\bar{l}$ comes from the
decay of $V$ on-shell.  Parity-violating asymmetries were defined and
presented at the parton level and for the hadronic processes $pp$ or
$p\bar{p}\rightarrow V$ + 2 jets or $l\bar{l}$ + 2 jets,
see Fig.~(\ref{cewi-had}).  These
asymmetries are independent of the polarizations of all particles
involved, and do not require that the flavors of the jet partons be
measured.  They are generally of order 0.01 pb at energies $\sqrt{s}
\gtrsim 1$ TeV.

\begin{figure}[h]
\centering{\includegraphics[width=8cm,clip]{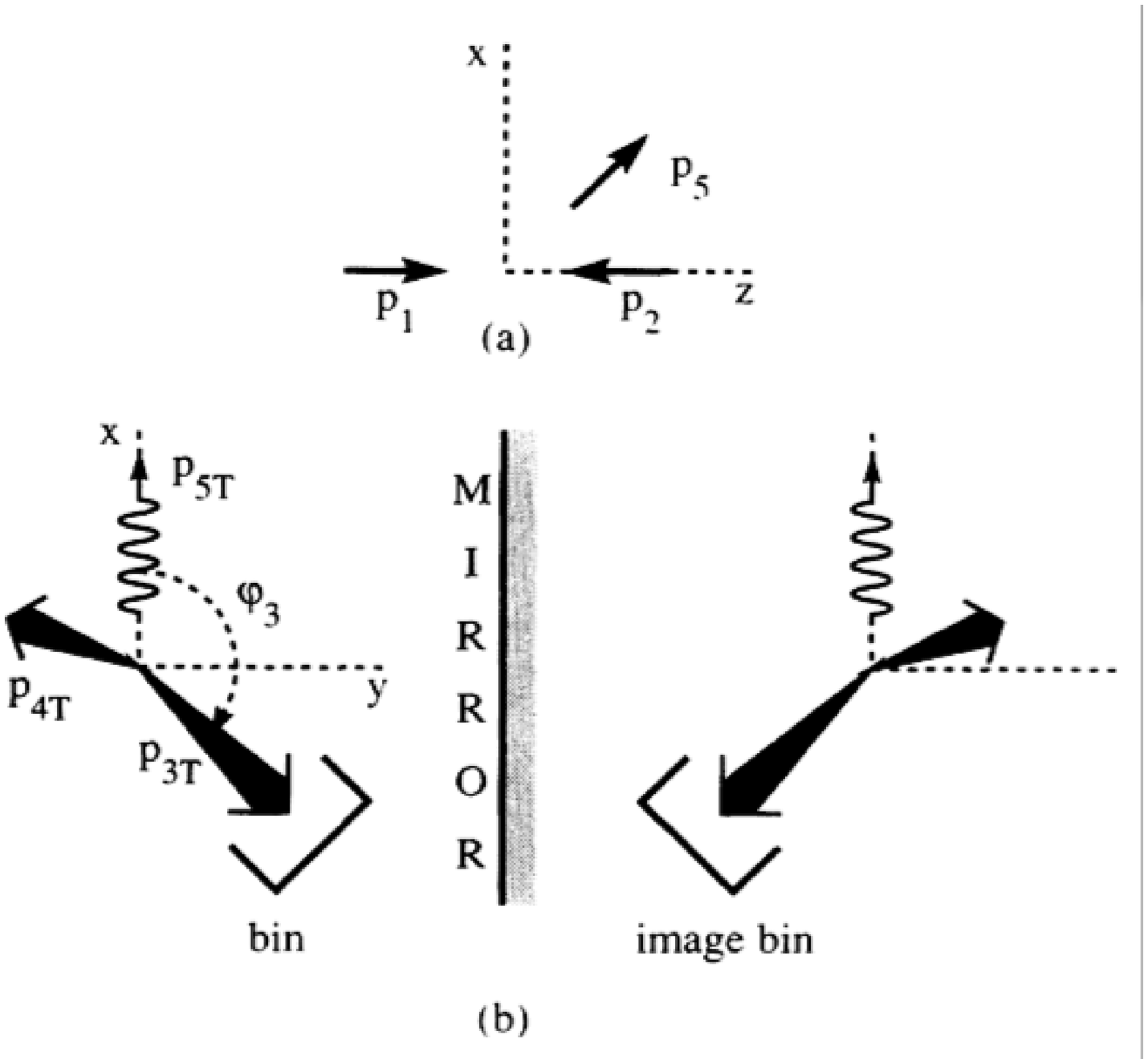}}
\vspace*{8pt}
\caption{
(a) Momenta in the $x{-}z$ plane.  $p_1$ is the momentum of the quark
(parton-level asymmetry), the proton ($p\bar p$ collisions), or one of
the protons chosen arbitrarily ($pp$ collisions).  $p_5$ is the
momentum of $V=W^\pm,Z^0$ (3-particle final state) or of the lepton
(4-particle final state).  A right-handed coordinate system is defined
such that $p_5^x>0$ and $p_5^y=0$. (b) Momenta in the $x{-}y$ plane.
$p_3$ and $p_4$ are the momenta of the two jet partons.  Parity is
violated if the event shown on the left and its mirror image shown on
the right occur with different probabilities.
}
\label{cewi-mirror}
\end{figure}

A simple way of defining experimental
observables that are sensitive to this parity-odd character of the
interference terms is as follows: Imagine that the incident beams lie
in the plane of a mirror, as in Fig.~(\ref{cewi-mirror}).
If the incident beams are not polarized,
the initial state is invariant under reflection in this mirror.  We
will also assume that the spins of the final state particles are not
detected, i.e., particles are identified by their momenta and internal
quantum numbers only.  A parity-odd contribution to the cross section
can make the probabilities for observing a particular event (i.e., a
particular configuration of final state particles) and its mirror
image different from one another.  Since the events are continuously
distributed in phase space, the likelihood of finding an event and its
geometrical mirror image in any finite sample of events is vanishingly
small.  To decide experimentally whether or not there is an asymmetry
with respect to mirror reflection in the event sample, one must count
the number of events that fall in some region of phase space (which we
will call a ``bin'') and the number of events that fall in the mirror
image of this region (which we will the ``image bin'') and then decide
whether or not there is a statistically significant difference between
these two numbers.  This difference can be compared with a
``parity-violating asymmetry'' which we define as follows:
\begin{equation}
a^{\rm pv}({\rm bin}) = \int\limits_{\rm bin}\,{\rm d}\sigma \quad-\quad
  \int\limits_{\rm image\,bin}{\rm d}\sigma\;.
\label{apveq}
\end{equation}
The integral over the bin includes an implicit sum over all final
state quantum numbers that are not observed, i.e., color quantum
numbers and any spin or flavor quantum numbers that cannot be
experimentally measured.

\begin{figure}[th]
\centering{\includegraphics[width=10cm,clip]{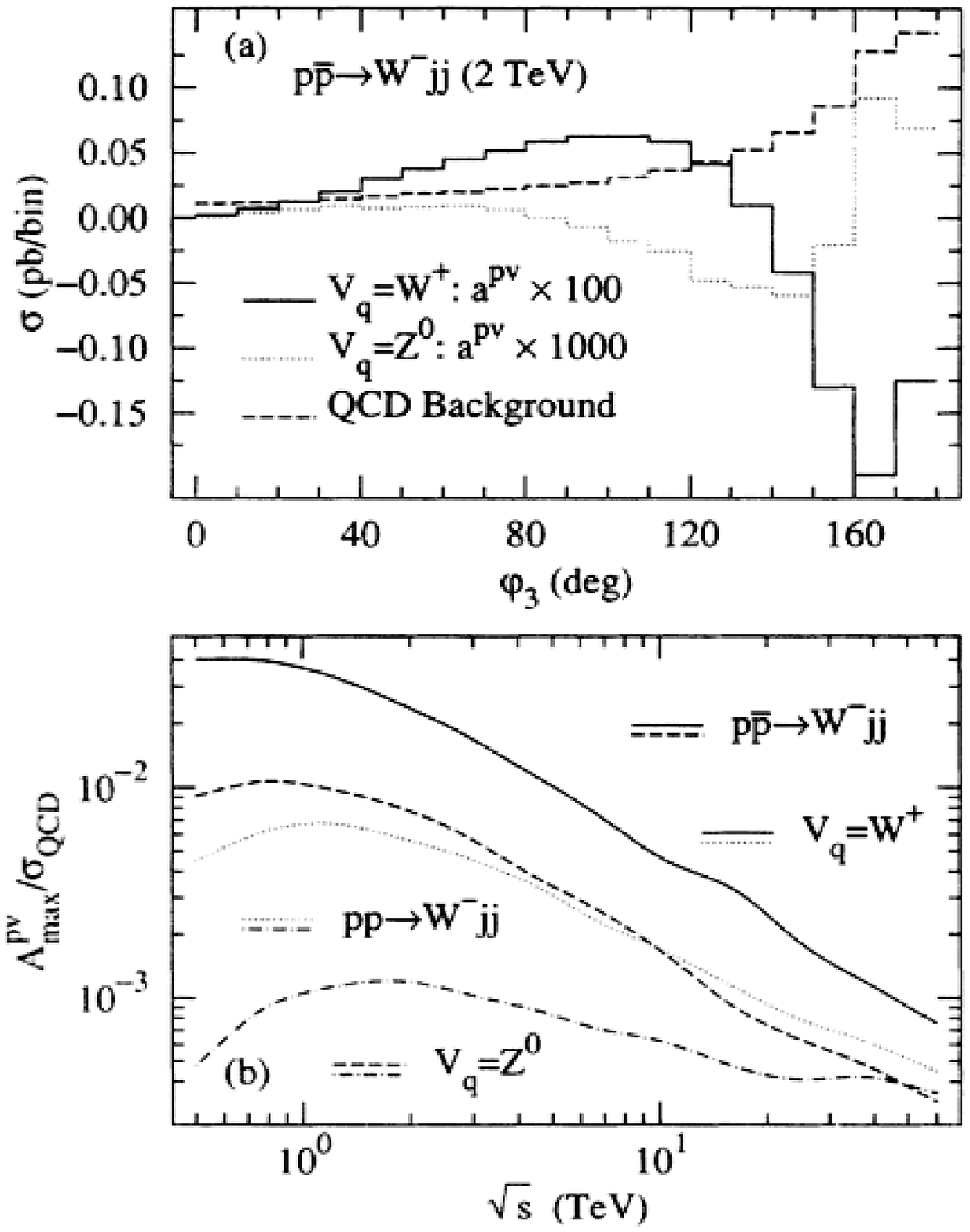}}
\vspace*{8pt}
\caption{
Comparison of signal to background in $W^-$ + 2 jet production.
(a) Binned cross sections with 200 GeV $<$ $\protect\sqrt{s_{12}}$
$<$ 600 GeV. The parity-violating asymmetries have been multiplied
by the factors (100 and 1000) indicated to show them on the same scale
as the QCD background.
(b) Variation of the signal to background
ratio in $pp$ and $p\bar p$ collisions as a function of the colliding
beam energy.
}
\label{cewi-signal}
\end{figure}

A set of bins will have to be judiciously chosen to maximize the
observed effects of parity violation, i.e., to yield the largest
cumulative asymmetry.  Thus we also define a cumulative
parity-violating asymmetry as follows:
\begin{equation}
A^{\rm pv} = \sum_{\rm bins} \mid a^{\rm pv}({\rm bin})\mid\;,
\label{Aeq}
\end{equation}
where the sum is taken over bins that do not overlap with one another.
A theoretical upper bound on $A^{\rm pv}$ is given by
\begin{eqnarray}
A^{\rm pv}_{\max} &=& \sum_{\rm bins} \lim_{\rm bin\,size\rightarrow 0}
 \mid a^{\rm pv}({\rm bin})\mid\\
 &=& \frac{1}{2}\int\limits_{\rm <cuts>} {\rm d}\Omega\;\left|
  \left(\frac{{\rm d}\sigma}{{\rm d}\Omega}\right)
  -\left(\frac{{\rm d}\sigma}{{\rm d}\Omega}
  \right)_{\rm mirror\,image} \right| \;.
\label{Amax2eq}
\end{eqnarray}

These asymmetries might be observable in $pp$ and $p\bar{p}$
collisions above the threshold for production of pairs of electroweak
bosons, i.e., at center of mass energies in the TeV range and these
asymmetries are comparable in magnitude to the pair-production cross
sections.
Fig.~(\ref{cewi-signal}) compares the parity-violating signal with
the QCD background\cite{Ellis:1985jg} in $W$ + 2 jet production.
Unlike the $W$-pair production cross section which is parity
conserving, the interference contribution is parity-violating, and
this might make it easier to observe above a rather formidable QCD
background of electroweak boson + 2 jet events.
Actual observation of
these effects will require somewhat higher integrated luminosities
than are currently available for example at the Tevatron at Fermilab.
It remains to be seen whether these subtle Standard Model predictions
will be observable at a high-energy multi-TeV hadron collider such as
the LHC.  Analogous parity-violating signatures may also be observable
in electron-positron annihilations\cite{Raina:2002uc} at the ILC.

\section{Electroweak Boson Production at Large Transverse Momentum}

The production of electroweak bosons at large transverse momentum is
one of the most important processes at current and future hadron
colliders.
The QCD-improved parton model predicts that if the intrinsic
energy scale involved is sufficiently large, the inclusive
cross section for the process
\begin{equation}
h_1(P_1)+h_2(P_2) \rightarrow V(Q) + X\;,
\end{equation}
where $h_i$, $i=1,2$ are unpolarized hadrons with momenta $P_i$,
can be reliably computed
using the following approximate factorized form:
\begin{equation}
E_Q\frac{{\rm d}\sigma}{{\rm d}^3Q}= \sum_{a_1,a_2}\int_0^1{\rm d}
x_2{\rm d} x_1
 f_{a_1}^{h_1}(x_1,M^2)f_{a_2}^{h_2}(x_2,M^2)E_Q
 \frac{{\rm d}\sigma^{a_1a_2}}{{\rm d}^3Q}
 (x_1P_1,x_2P_2,M^2)\;.
\end{equation}
Here $E_Q\equiv Q^0$, $a$ and $b$ stand for quarks, antiquarks or gluons,
$f_a^h(x,M^2)$ is the probability density for finding parton $a$
with momentum fraction $x$
in hadron $h$ if it is probed at scale $M^2$, and
$\sigma^{ab}(p_1,p_2,M^2)$ is the perturbative cross section for the process
\begin{equation}
a(p_1)+b(p_2)\rightarrow V(Q)+X\;,
\end{equation}
from which collinear singularities arising from radiation off massless
partons have been factorized out at scale $M^2$ and implicitly
included in the scale-dependent parton densities $f_a^h(x,M^2)$.

\begin{figure}[th]
\centering{\includegraphics[width=10cm,clip]{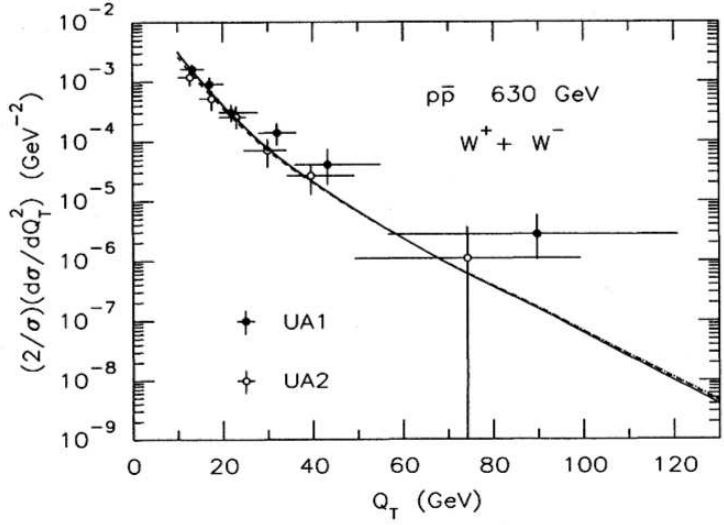}}
\vspace*{8pt}
\caption{
$W$ production in $p\bar p$ collisions at 630 GeV.
The solid, dashed, and dotted lines represent the Full (MRSB),
Full (MRSE), and NS (MRSB) predictions.  The corresponding total
cross sections $\sigma$ are 5.8, 5.1, and 3.5 nb, respectively.}
\label{ua12}
\end{figure}

A complete analytical calculation of the next-to-leading order QCD
radiative corrections to the inclusive cross sections $parton + parton
\rightarrow V + X$, where $V$ is an on-shell $W^\pm$ or $Z^0$ with
transverse momentum $Q_T$ of order $M_W$, or a massive virtual photon
with $Q_T$ of order of its invariant mass, was done by Gonsalves,
Paw{\l}owski and Wai\cite{Gonsalves:1989ar} and by Arnold and
Reno\cite{Arnold:1988dp}.  This work completed an earlier calculation
by Ellis, Martinelli and Petronzio\cite{Ellis:1981hk} of the
non-singlet (NS) contributions to lepton-pair production at large
$Q_T$.  Numerical predictions for $W$, $Z$ and $\gamma^*$ production
at collider energies were computed, and the dependence of the
radiative corrections on the choice of renormalization and
factorization scales was studied.  Results\cite{Gonsalves:1990ae} on
varying the renormalization and factorization scales independently
were used to compare the FAC\cite{Grunberg:1980ja} and
PMS\cite{Stevenson:1981vj} prescriptions.  These results showed
that the QCD-improved parton model can be used to make firm and
reliable predictions for electroweak boson production at large $Q_T$.

In Fig.~(\ref{ua12}), the theoretical predictions are compared with
measurements\cite{Albajar:1987aj,Ansari:1988sv} of the $W$ transverse
momentum distributions by the UA1 and UA2 collaborations at the CERN
Sp$\bar{\rm p}$S collider.  It is clear that data are consistent with
the QCD predictions, but that they are not accurate enough to
discriminate between the MRSB and MRSE
distributions\cite{Martin:1988aj} or between the singlet and
non-singlet theoretical predictions.  There is a tantalizing, but
statistically insignificant, hint in the data that the experimental
cross section might be larger than the theoretical prediction at large
values of $Q_T$.

\begin{figure}[th]
\centering{\includegraphics[width=10cm,clip]{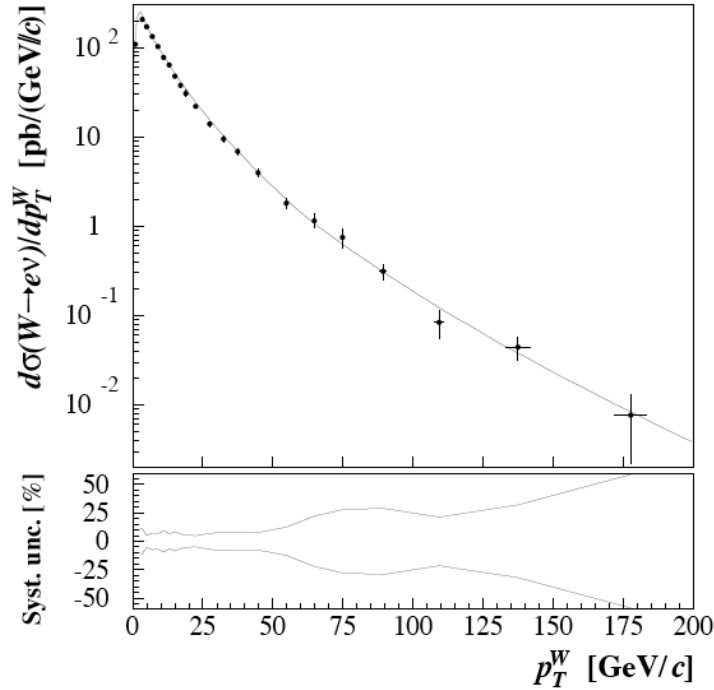}}
\vspace*{8pt}
\caption{
$W$ production in $p\bar p$ collisions at 1.8 TeV measured by the
D0 Collaboration\protect\cite{Abbott:2000xv}.}
\label{dzero}
\end{figure}

Fig.~(\ref{dzero}) shows a comparison of theoretical predictions with
experimental measurements by the D0 Collaboration.
The theoretical results in this figure have been improved at small $Q_T$
by resumming Sudakov logarithms of $Q_T^2/M_W^2$.
The data are completely consistent with the predictions and
there is no evidence in the tail of the distribution for new physics
beyond the Standard Model at large $Q_T$.

\subsection{Large logarithmic soft-gluon corrections}

Predictions for the $W$ transverse momentum distribution at
next-to-leading order (NLO) in $\alpha_s$ and including the
next-to-next-to-leading order (NNLO) soft-gluon corrections at LHC
energies were presented recently by Gonsalves, Kidonakis and
Sabio-Vera\cite{Gonsalves:2005ng}.

Let $s_2$ be the invariant mass of the final state partons
recoiling against the $W$ boson.
In general, the partonic cross section $\hat{\sigma}$
includes distributions with respect
to $s_2$ at $n$-th order in the QCD coupling $\alpha_s$ of the type
\begin{equation}
\left[\frac{\ln^{m}(s_2/Q_T^2)}{s_2} \right]_+, \hspace{10mm} m\le 2n-1\, ,
\label{zplus}
\end{equation}
defined by their integral with any smooth function $f$ by
\begin{eqnarray}
\int_0^{s_{2 \, \rm max}} {\rm d}s_2 \, f(s_2) \left[\frac{\ln^m(s_2/Q_T^2)}
{s_2}\right]_{+} &\equiv&
\int_0^{s_{2\, \rm max}} {\rm d}s_2 \frac{\ln^m(s_2/Q_T^2)}{s_2} [f(s_2) - f(0)]
\nonumber \\ &&
{}+\frac{1}{m+1} \ln^{m+1}\left(\frac{s_{2\, \rm max}}{Q_T^2}\right) f(0) \, .
\label{splus}
\end{eqnarray}
These ``plus'' distributions are the remnants of cancellations between
real and virtual contributions to the cross section.
Let us make use of the terminology that at $n$-th order in $\alpha_s$
the leading logarithms (LL)
are those with $m=2n-1$ in Eq. (\ref{zplus}),
the next-to-leading logarithms (NLL) are those with $m=2n-2$,
the next-to-next-to-leading logarithms (NNLL) are those with $m=2n-3$,
and the next-to-next-to-next-to-leading logarithms (NNNLL)
are those with $m=2n-4$.
The symbol ``NNLO-NNNLL'' means that soft-gluon contributions
through NNNLL to the NNLO corrections have been included and added to the
complete NLO result.

\begin{figure}[th]
\centering{\includegraphics[width=10cm,clip]{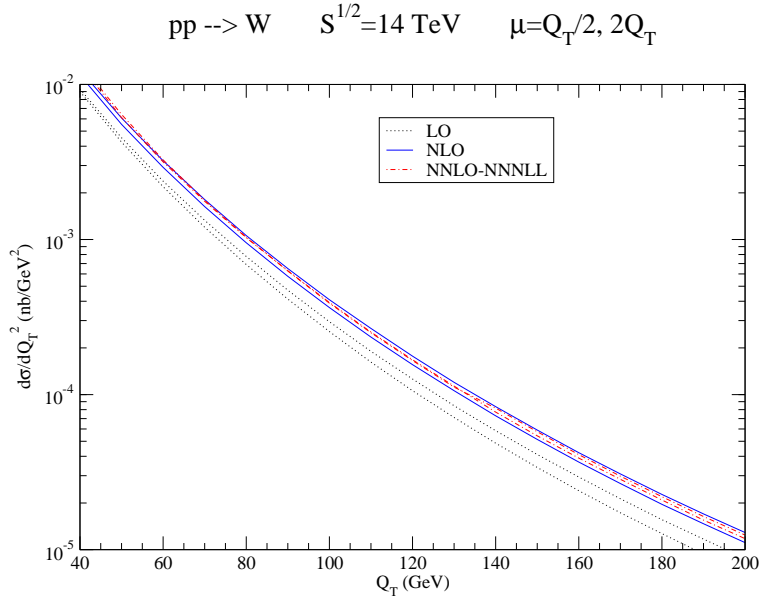}}
\vspace*{8pt}
\caption{The differential cross section,
$d\sigma/dQ_T^2$, for $W$ production in $pp$ collisions
at the LHC with $\sqrt{S}=14$ TeV and
$\mu=\mu_F=\mu_R=Q_T/2$ or $2Q_T$.
Shown are the LO, NLO,
and NNLO-NNNLL results using the MRST 2002 LO, NLO and NNLO
parton distributions\protect\cite{Martin:2002aw}. The upper lines
are with $\mu=Q_T/2$, the lower lines with $\mu=2 Q_T$.}
\label{soft-gluon}
\end{figure}

Fig.~(\ref{soft-gluon}) shows the differential cross section
$d\sigma/dQ_T^2$ at high $Q_T$ with $\sqrt{S}=14$ TeV for the two
scale values, $Q_T/2$ and $2Q_T$, often used to display the
uncertainty due to scale variation.  Note that while the variation of
the Born cross section is significant and the variation at NLO is
similar to LO, at NNLO-NNNLL it is very small. In fact the two
NNLO-NNNLL curves lie very close to or on top of each other.  With
this calculation, reliable predictions of perturbative QCD, including
soft-gluon effects, for the $W$ transverse momentum distribution are
available and await comparison with experimental measurements at the
LHC in the near future.  Any deviation from these predictions would be
an unambiguous signal of new physics beyond the Standard Model.

It is obviously worthwhile to compute the complete NNLO corrections to
the transverse momentum distribution.  This would further reduce the
scale dependence of the predictions, and it would also provide
information on soft and logarithms at ${\cal O}(\alpha_s^4)$.  Fixed
order QCD calculations for analogous processes in $e^+e^-$
annihilation have been completed over the past few years, see e.g.
Gehrmann-De Ridder et al.\cite{GehrmannDeRidder:2008ug}, and
references therein.  An ${\cal O}(\alpha_s^3)$ calculation at finite
$Q_T$ would require the two-loop corrections to subprocesses such as
$q\bar q\rightarrow Wg$, one-loop corrections to subprocesses such as
$q\bar q\rightarrow Wgg$, and tree-level amplitudes such as $q\bar
q\rightarrow Wggg$.

\section{Heavy Quark Triangle Diagram Contributions at High Energies}

Because the masses of the $W$ and $Z$ bosons are large compared with
the masses of the quark and gluon constituents in the colliding
hadrons, it is generally a good approximation to assume that the
quarks are massless in computing the production cross sections for
these electroweak bosons at high energies.
An exception, of course is the top quark, whose mass $m_t=174$~GeV, is
approximately twice as large as that of the $W$ or $Z$.

The top quark can contribute to production amplitudes in three ways:
as an initial state partonic constituent of one of the colliding
hadrons, as an on-shell particle in the final state, and as a virtual
particle.
The density of top quarks in the colliding hadrons can safely be
neglected.
The production of real top quarks can safely be treated in a threshold
approximation with the mass taken to be infinite below threshold and
zero above.
Top quarks in loops pose a delicate problem because the Standard Model
is renormalizable only if the left-handed quarks occur in SU(2)$_L$
doublets.
This requirement is imposed by the well-known triangle anomaly
(which was discovered in QED by Adler\cite{Adler:1969gk} and
Bell and Jackiw\cite{Bell:1969ts})
in the effective
coupling of the axial-vector current to two gluons: if the top quark
is omitted from the loop the amplitude is divergent in a
non-renormalizable way; if the top quark is included and assumed
massless the top and bottom contributions cancel; and if the top
quark is included with a finite mass the contribution grows as the
logarithm of the ratio of top to bottom masses.

\begin{figure}[th]
\centering{
\includegraphics[width=5.5cm,clip]{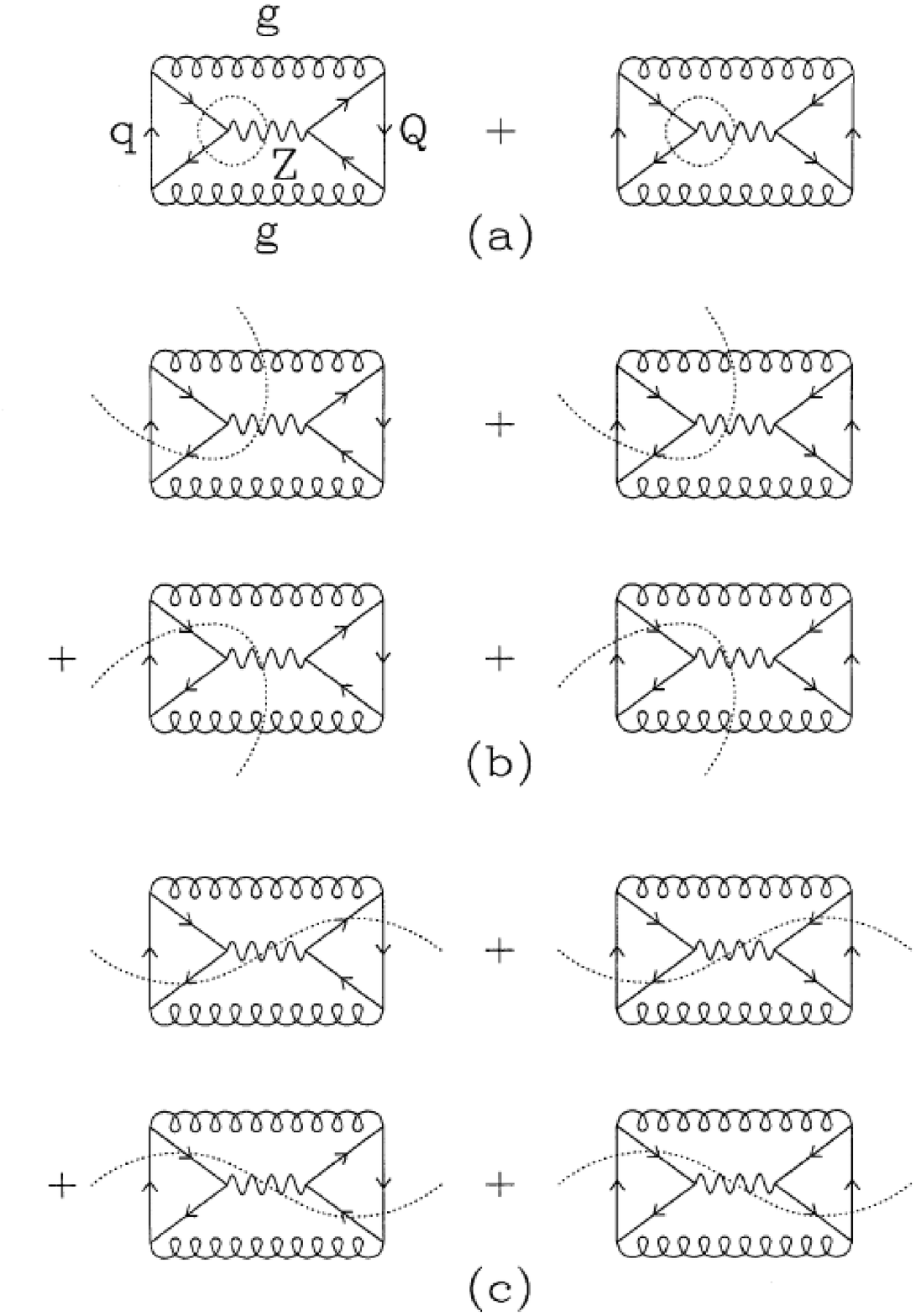}
\includegraphics[width=6.5cm,clip]{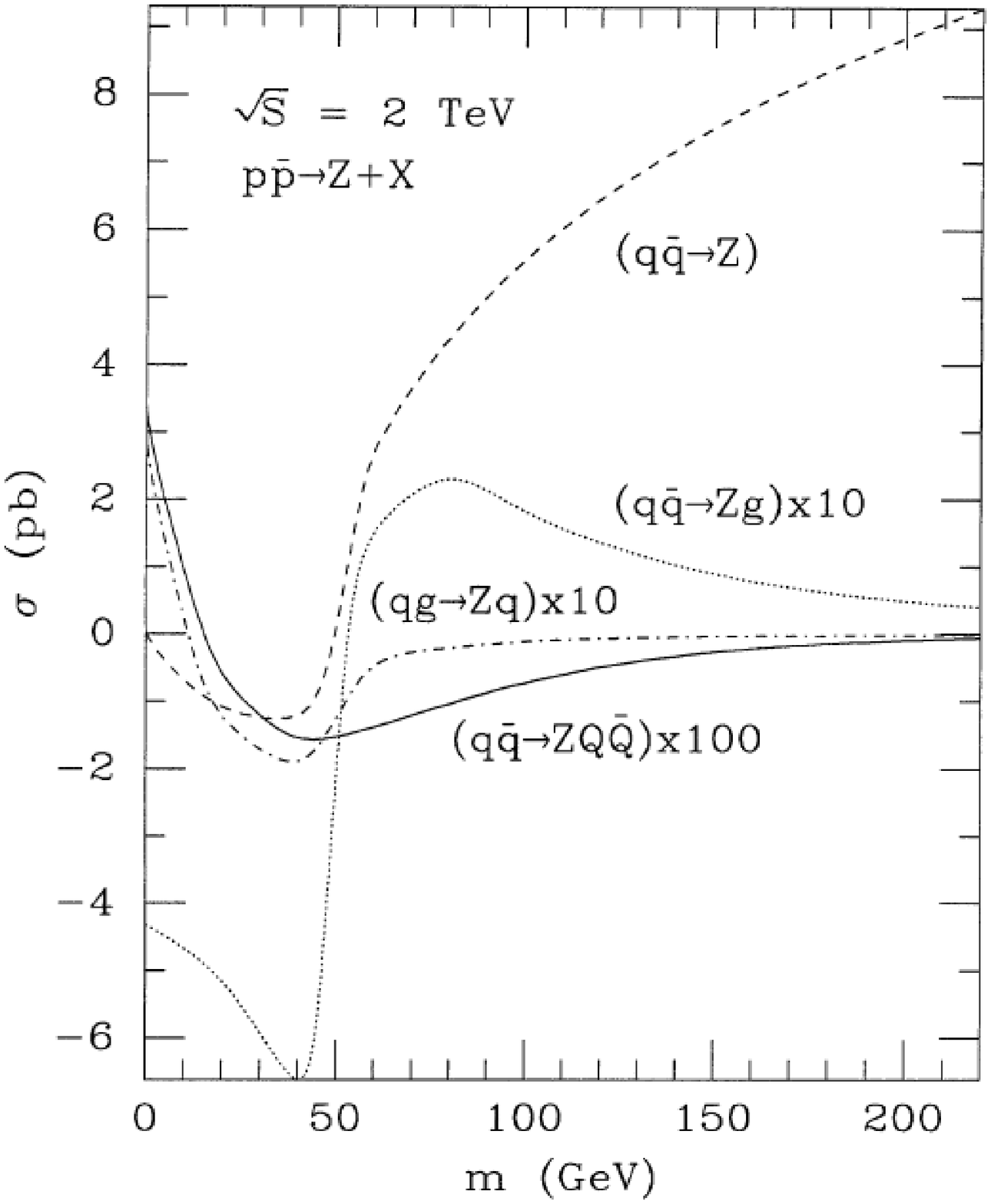}
}
\vspace*{8pt}
\caption{
Left panel: Diagrams which contribute to the total cross section for
producing a $Z$ boson in hadron collisions.  $Q$ and $q$ represent
heavy and massless quarks, respectively, and $g$ a gluon.  Propagators
cut by a dotted line are taken to be on shell in the initial or final
state.  Right panel: Contributions from heavy-quark triangle diagrams
to the total cross section for $Z$ production in $p\bar p$ collisions
at 2 TeV.  $m$ is the mass of the heavy quark, which is assumed to have
weak isospin ${+}\frac{1}{2}$.  A quark with weak isospin ${-}\frac{1}{2}$
would contribute with opposite sign.  The total cross section at 2 TeV is
approximately 5 nb.
}
\label{triang92-fig}
\end{figure}

The contributions in second order QCD of the diagrams in the left
panel of Fig.~(\ref{triang92-fig}) involving a heavy-quark triangle
with vector, vector and axial-vector vertices were studied by
Gonsalves, Hung and Paw{\l}owski\cite{Gonsalves:1991qn}.  The
contributions from the process $q\bar q\rightarrow Zgg$ ((a) in
Fig.~(\ref{triang92-fig})) had been analyzed earlier by Dicus and
Willenbrock\cite{Dicus:1985wx}, who concluded that the contributions
were numerically small at Tevatron energies. The right panel of
Fig.~(\ref{triang92-fig}) shows the numerical magnitudes of the
various subprocess contributions at Tevatron energies.

Rijken and van Neerven\cite{Rijken:1995gi} subsequently computed all
heavy quark contributions to the $W$ and $Z$ total cross sections,
neglecting only diagrams with heavy quarks in the initial state.
They found that the contributions from the top quark were negligible
at Tevatron energies, but would contribute significantly at LHC
energies with magnitude comparable to the ${\cal O}(\alpha_s^2)$
QCD corrections.

\begin{figure}[th]
\centering{\includegraphics[width=10cm,clip]{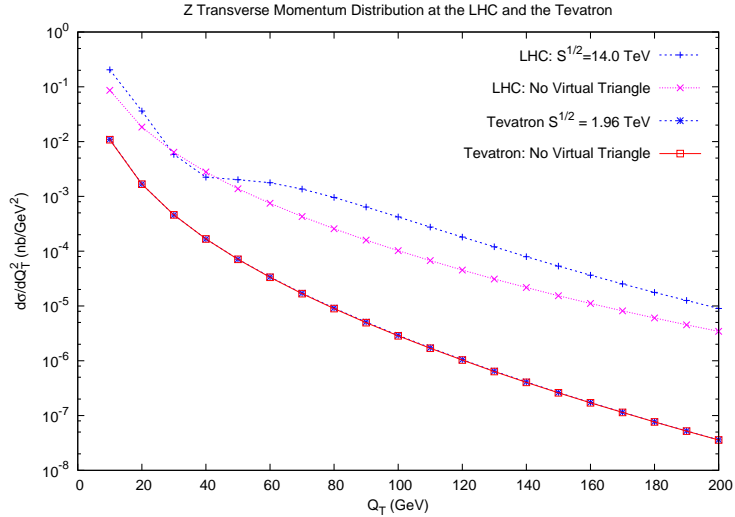}}
\vspace*{8pt}
\caption{
$Z$ production at the LHC and the Tevatron.
The data points represent NLO predictions computed using MRST 2002
NLO parton distributions\protect\cite{Martin:2002aw}.
Note the change in shape of the distribution
at LHC energies (upper curves) compared with Tevatron energies (lower curves)
when the anomaly diagrams are included.  The anomalous contributions are
negligibly small at the Tevatron.}
\label{triang08-fig}
\end{figure}

It is important to evaluate the top-quark-mass contributions to other
measurable cross sections at LHC energies, in particular the $W$ and
$Z$ transverse momentum distributions.  Fig.~(\ref{triang08-fig})
shows the $Z$ transverse-momentum distribution at Tevatron and LHC
energies including quark-triangle diagrams shown in
Fig.~(\ref{triang92-fig}).  Quark masses are taken into account using
a simple threshold prescription\cite{Gonsalves:1989ar} by inserting
a step function $\theta(Q_T^2-4m_f^2)$ into sums over quark flavors $f$
in the subprocess cross sections.  There is a significant change in
the shape of the distribution at LHC energies due specifically to the
virtual triangle-anomaly diagrams.

Because the triangle-anomaly contributions are comparable in magnitude
to the ${\cal O}(\alpha_s^2)$ QCD radiative corrections, it is
essential to include the top and bottom quark masses in the virtual
triangle diagrams.  A more precise calculation of the triangle-anomaly
contributions to the $Z$ transverse momentum distribution at LHC
energies will be published elsewhere.

\section*{Acknowledgements}

I would like to acknowledge my collaborators, W.~Celmaster, C.W.~Choi,
R.K.~Ellis, C.M.~Hung, N.~Kidonakis, J.~Paw{\l}owski, S.~Raina,
A.~Sabio-Vera, and C.F.~Wai, without whose collaboration many of the
calculations described here would not have been possible.  I am
grateful to the National Science Foundation for support over many
years, most recently through NSF-PHY-92-12177, as well as support
provided by the Center for Computational Research at the University at
Buffalo.


\begin{thebibliography}{99}

%\cite{Jain:1997pc}
\bibitem{Jain:1997pc}
  P.~L.~Jain, G.~Singh and R.~Gonsalves,
  %``Phase transition in gold nuclei at 10.6-A-GeV,''
  Mod.\ Phys.\ Lett.\  A {\bf 12} (1997) 1101.
  %%CITATION = MPLAE,A12,1101;%%

%\cite{Celmaster:1979dm}
\bibitem{Celmaster:1979dm}
  W.~Celmaster and R.~J.~Gonsalves,
  %``QCD Perturbation Expansions In A Coupling Constant Renormalized By Momentum
  %Space Subtraction,''
  Phys.\ Rev.\ Lett.\  {\bf 42} (1979) 1435.
  %%CITATION = PRLTA,42,1435;%%

%\cite{Celmaster:1979km}
\bibitem{Celmaster:1979km}
  W.~Celmaster and R.~J.~Gonsalves,
  %``The Renormalization Prescription Dependence Of The QCD Coupling Constant,''
  Phys.\ Rev.\  D {\bf 20} (1979) 1420.
  %%CITATION = PHRVA,D20,1420;%%

%\cite{EllisStirlingWebber}
\bibitem{EllisStirlingWebber}
  R. K. Ellis, W. J. Stirling and B. R. Webber, {\it QCD and Collider Physics}
  (Cambridge University Press, Cambridge, 1996).

%\cite{Bardeen:1978yd}
\bibitem{Bardeen:1978yd}
  W.~A.~Bardeen, A.~J.~Buras, D.~W.~Duke and T.~Muta,
  %``Deep Inelastic Scattering Beyond The Leading Order In Asymptotically Free
  %Gauge Theories,''
  Phys.\ Rev.\  D {\bf 18} (1978) 3998.
  %%CITATION = PHRVA,D18,3998;%%

%\cite{'t Hooft:1972fi}
\bibitem{'t Hooft:1972fi}
  G.~'t Hooft and M.~J.~G.~Veltman,
  %``Regularization And Renormalization Of Gauge Fields,''
  Nucl.\ Phys.\  B {\bf 44}, 189 (1972).
  %%CITATION = NUPHA,B44,189;%%

%\cite{Brodsky:1982gc}
\bibitem{Brodsky:1982gc}
  S.~J.~Brodsky, G.~P.~Lepage and P.~B.~Mackenzie,
  %``On The Elimination Of Scale Ambiguities In Perturbative Quantum
  %Chromodynamics,''
  Phys.\ Rev.\  D {\bf 28} (1983) 228.
  %%CITATION = PHRVA,D28,228;%%

%\cite{Grunberg:1980ja}
\bibitem{Grunberg:1980ja}
  G.~Grunberg,
  %``Renormalization Group Improved Perturbative QCD,''
  Phys.\ Lett.\  B {\bf 95} (1980) 70;
  Erratum {\it ibid}.\  B {\bf 110} (1982) 501.
  %%CITATION = PHLTA,B95,70;%%

%\cite{Stevenson:1981vj}
\bibitem{Stevenson:1981vj}
  P.~M.~Stevenson,
  %``Optimized Perturbation Theory,''
  Phys.\ Rev.\  D {\bf 23} (1981) 2916.
  %%CITATION = PHRVA,D23,2916;%%

%\cite{GellMann:1954fq}
\bibitem{GellMann:1954fq}
  M.~Gell-Mann and F.~E.~Low,
  %``Quantum electrodynamics at small distances,''
  Phys.\ Rev.\  {\bf 95} (1954) 1300.
  %%CITATION = PHRVA,95,1300;%%

%\cite{Stueckelberg:1953dz}
\bibitem{Stueckelberg:1953dz}
  E.~C.~G.~Stueckelberg and A.~Petermann,
  %``Normalization of constants in the quanta theory,''
  Helv.\ Phys.\ Acta {\bf 26} (1953) 499.
  %%CITATION = HPACA,26,499;%%

%\cite{Gross:1973id}
\bibitem{Gross:1973id}
  D.~J.~Gross and F.~Wilczek,
  %``ULTRAVIOLET BEHAVIOR OF NON-ABELIAN GAUGE THEORIES,''
  Phys.\ Rev.\ Lett.\  {\bf 30} (1973) 1343.
  %%CITATION = PRLTA,30,1343;%%

%\cite{Politzer:1973fx}
\bibitem{Politzer:1973fx}
  H.~D.~Politzer,
  %``RELIABLE PERTURBATIVE RESULTS FOR STRONG INTERACTIONS?,''
  Phys.\ Rev.\ Lett.\  {\bf 30} (1973) 1346.
  %%CITATION = PRLTA,30,1346;%%

%\cite{Hasenfratz:1980kn}
\bibitem{Hasenfratz:1980kn}
  A.~Hasenfratz and P.~Hasenfratz,
  %``The Connection Between The Lambda Parameters Of Lattice And Continuum
  %QCD,''
  Phys.\ Lett.\  B {\bf 93} (1980) 165.
  %%CITATION = PHLTA,B93,165;%%

%\cite{Dashen:1980vm}
\bibitem{Dashen:1980vm}
  R.~F.~Dashen and D.~J.~Gross,
  %``The Relationship Between Lattice And Continuum Definitions Of The Gauge
  %Theory Coupling,''
  Phys.\ Rev.\  D {\bf 23} (1981) 2340.
  %%CITATION = PHRVA,D23,2340;%%

%\cite{Yao:2006px}
\bibitem{Yao:2006px}
  W.~M.~Yao {\it et al.}  [Particle Data Group],
  %``Review of particle physics,''
  J.\ Phys.\ G {\bf 33} (2006) 1.
  %%CITATION = JPHGB,G33,1;%%

%\cite{Schwinger:1948yk}
\bibitem{Schwinger:1948yk}
  J.~S.~Schwinger,
  %``Quantum electrodynamics. I: A covariant formulation,''
  Phys.\ Rev.\  {\bf 74} (1948) 1439.
  %%CITATION = PHRVA,74,1439;%%

%\cite{Dine:1979qh}
\bibitem{Dine:1979qh}
  M.~Dine and J.~R.~Sapirstein,
  %``Higher Order QCD Corrections In E+ E- Annihilation,''
  Phys.\ Rev.\ Lett.\  {\bf 43} (1979) 668.
  %%CITATION = PRLTA,43,668;%%

%\cite{Celmaster:1979xr}
\bibitem{Celmaster:1979xr}
  W.~Celmaster and R.~J.~Gonsalves,
  %``An Analytic Calculation Of Higher Order Quantum Chromodynamic Corrections
  %In E+ E- Annihilation,''
  Phys.\ Rev.\ Lett.\  {\bf 44} (1980) 560.
  %%CITATION = PRLTA,44,560;%%

%\cite{Celmaster:1980ji}
\bibitem{Celmaster:1980ji}
  W.~Celmaster and R.~J.~Gonsalves,
  %``Fourth Order QCD Contributions To The E+ E- Annihilation Cross-Section,''
  Phys.\ Rev.\  D {\bf 21} (1980) 3112.
  %%CITATION = PHRVA,D21,3112;%%

%\cite{Chetyrkin:1979bj}
\bibitem{Chetyrkin:1979bj}
  K.~G.~Chetyrkin, A.~L.~Kataev and F.~V.~Tkachov,
  %``Higher Order Corrections To Sigma-T (E+ E- $\to$ Hadrons) In Quantum
  %Chromodynamics,''
  Phys.\ Lett.\  B {\bf 85} (1979) 277.
  %%CITATION = PHLTA,B85,277;%%

%\cite{DeRafael:1974iv}
\bibitem{DeRafael:1974iv}
  E.~De Rafael and J.~L.~Rosner,
  %``Short-Distance Behavior Of Quantum Electrodynamics And The Callan-Symanzik
  %Equation For The Photon Propagator,''
  Annals Phys.\  {\bf 82} (1974) 369.
  %%CITATION = APNYA,82,369;%%

%\cite{Abramowitz}
\bibitem{Abramowitz}
  M. Abramowitz and I. A. Stegun, {\it Handbook of Mathematical Functions}
  (Dover, New York, 1972)

%\cite{Levine:1974xh}
\bibitem{Levine:1974xh}
  M.~J.~Levine and R.~Roskies,
  %``Hyperspherical approach to quantum electrodynamics - sixth-order magnetic
  %moment,''
  Phys.\ Rev.\  D {\bf 9} (1974) 421.
  %%CITATION = PHRVA,D9,421;%%

%\cite{Vladimirov:1979zm}
\bibitem{Vladimirov:1979zm}
  A.~A.~Vladimirov,
  %``Method For Computing Renormalization Group Functions In Dimensional
  %Renormalization Scheme,''
  Theor.\ Math.\ Phys.\  {\bf 43} (1980) 417
  [Teor.\ Mat.\ Fiz.\  {\bf 43} (1980) 210].
  %%CITATION = TMFZA,43,210;%%

%\cite{Gorishnii:1988bc}
\bibitem{Gorishnii:1988bc}
  S.~G.~Gorishnii, A.~L.~Kataev and S.~A.~Larin,
  %``Next-To-Leading O(alpha-s**3) QCD Correction to Sigma-t (e+ e- $\to$
  %Hadrons): Analytical Calculation and Estimation of the Parameter Lambda
  %(MS),''
  Phys.\ Lett.\  B {\bf 212} (1988) 238.
  %%CITATION = PHLTA,B212,238;%%

%\cite{Surguladze:1990tg}
\bibitem{Surguladze:1990tg}
  L.~R.~Surguladze and M.~A.~Samuel,
  %``Total Hadronic Cross-Section In E+ E- Annihilation At The Four Loop Level
  %Of Perturbative QCD,''
  Phys.\ Rev.\ Lett.\  {\bf 66} (1991) 560;
  Erratum {\it ibid}.\  {\bf 66} (1991) 2416.
  %%CITATION = PRLTA,66,560;%%

%\cite{Baikov:2006nb}
\bibitem{Baikov:2006nb}
  P.~A.~Baikov, K.~G.~Chetyrkin and J.~H.~Kuhn,
  %``Multiloop calculations: Towards R at order alpha(s)**4,''
  Nucl.\ Phys.\ Proc.\ Suppl.\  {\bf 157} (2006) 27
  [arXiv:hep-ph/0602126].
  %%CITATION = NUPHZ,157,27;%%

%\cite{Collins:1989bt}
\bibitem{Collins:1989bt}
  J.~C.~Collins,
  %``Sudakov form factors,''
  Adv.\ Ser.\ Direct.\ High Energy Phys.\  {\bf 5} (1989) 573
  [arXiv:hep-ph/0312336].
  %%CITATION = 00319,5,573;%%

%\cite{Gonsalves:1983nq}
\bibitem{Gonsalves:1983nq}
  R.~J.~Gonsalves,
  %``Dimensionally Regularized Two Loop On-Shell Quark Form-Factor,''
  Phys.\ Rev.\  D {\bf 28} (1983) 1542.
  %%CITATION = PHRVA,D28,1542;%%

%\cite{van Neerven:1985xr}
\bibitem{van Neerven:1985xr}
  W.~L.~van Neerven,
  %``Dimensional Regularization Of Mass And Infrared Singularities In Two Loop
  %On-Shell Vertex Functions,''
  Nucl.\ Phys.\  B {\bf 268} (1986) 453.
  %%CITATION = NUPHA,B268,453;%%

%\cite{Kramer:1986sr}
\bibitem{Kramer:1986sr}
  G.~Kramer and B.~Lampe,
  %``INTEGRALS FOR TWO LOOP CALCULATIONS IN MASSLESS QCD,''
  J.\ Math.\ Phys.\  {\bf 28} (1987) 945.
  %%CITATION = JMAPA,28,945;%%

%\cite{Gehrmann:2005pd}
\bibitem{Gehrmann:2005pd}
  T.~Gehrmann, T.~Huber and D.~Maitre,
  %``Two-loop quark and gluon form factors in dimensional regularisation,''
  Phys.\ Lett.\  B {\bf 622} (2005) 295
  [arXiv:hep-ph/0507061].
  %%CITATION = PHLTA,B622,295;%%

%\cite{Hamberg:1990np}
\bibitem{Hamberg:1990np}
  R.~Hamberg, W.~L.~van Neerven and T.~Matsuura,
  %``A Complete calculation of the order alpha-s**2 correction to the Drell-Yan
  %K factor,''
  Nucl.\ Phys.\  B {\bf 359} (1991) 343
  Erratum, {\it ibid}.\  B {\bf 644} (2002) 403.
  %%CITATION = NUPHA,B359,343;%%

%\cite{Harlander:2002wh}
\bibitem{Harlander:2002wh}
  R.~V.~Harlander and W.~B.~Kilgore,
  %``Next-to-next-to-leading order Higgs production at hadron colliders,''
  Phys.\ Rev.\ Lett.\  {\bf 88} (2002) 201801
  [arXiv:hep-ph/0201206].
  %%CITATION = PRLTA,88,201801;%%

%\cite{Anastasiou:2002yz}
\bibitem{Anastasiou:2002yz}
  C.~Anastasiou and K.~Melnikov,
  %``Higgs boson production at hadron colliders in NNLO QCD,''
  Nucl.\ Phys.\  B {\bf 646}, 220 (2002)
  [arXiv:hep-ph/0207004].
  %%CITATION = NUPHA,B646,220;%%

%\cite{Anastasiou:2003ds}
\bibitem{Anastasiou:2003ds}
  C.~Anastasiou, L.~J.~Dixon, K.~Melnikov and F.~Petriello,
  %``High-precision QCD at hadron colliders: Electroweak gauge boson  rapidity
  %distributions at NNLO,''
  Phys.\ Rev.\  D {\bf 69} (2004) 094008
  [arXiv:hep-ph/0312266].
  %%CITATION = PHRVA,D69,094008;%%

%\cite{Ravindran:2003um}
\bibitem{Ravindran:2003um}
  V.~Ravindran, J.~Smith and W.~L.~van Neerven,
  %``NNLO corrections to the total cross section for Higgs boson production  in
  %hadron hadron collisions,''
  Nucl.\ Phys.\  B {\bf 665}, 325 (2003)
  [arXiv:hep-ph/0302135];
  %%CITATION = NUPHA,B665,325;%%
%
%\cite{Ravindran:2006bu}
%\bibitem{Ravindran:2006bu}
%  V.~Ravindran, J.~Smith and W.~L.~van Neerven,
  %``QCD threshold corrections to di-lepton and Higgs rapidity distributions
  %beyond N(2)LO,''
  Nucl.\ Phys.\  B {\bf 767}, 100 (2007)
  [arXiv:hep-ph/0608308].
  %%CITATION = NUPHA,B767,100;%%

%\cite{Gonsalves:1986af}
\bibitem{Gonsalves:1986af}
  R.~J.~Gonsalves,
  %``INVARIANT AMPLITUDES FOR q anti-q g FINAL STATES IN e+ e- ANNIHILATION,''
  Phys.\ Rev.\  D {\bf 34} (1986) 1316.
  %%CITATION = PHRVA,D34,1316;%%

%\cite{Mastrolia:2003yz}
\bibitem{Mastrolia:2003yz}
  P.~Mastrolia and E.~Remiddi,
  %``Two-loop form factors in QED,''
  Nucl.\ Phys.\  B {\bf 664} (2003) 341
  [arXiv:hep-ph/0302162].
  %%CITATION = NUPHA,B664,341;%%

%\cite{Bonciani:2003ai}
\bibitem{Bonciani:2003ai}
  R.~Bonciani, P.~Mastrolia and E.~Remiddi,
  %``QED vertex form factors at two loops,''
  Nucl.\ Phys.\  B {\bf 676} (2004) 399
  [arXiv:hep-ph/0307295].
  %%CITATION = NUPHA,B676,399;%%

%\cite{Bernreuther:2005gq}
\bibitem{Bernreuther:2005gq}
  W.~Bernreuther, R.~Bonciani, T.~Gehrmann, R.~Heinesch, T.~Leineweber, P.~Mastrolia and E.~Remiddi,
  %``QCD corrections to static heavy quark form factors,''
  Phys.\ Rev.\ Lett.\  {\bf 95} (2005) 261802
  [arXiv:hep-ph/0509341];
  %%CITATION = PRLTA,95,261802;%%
%
%\cite{Bernreuther:2005gw}
%\bibitem{Bernreuther:2005gw}
%  W.~Bernreuther, R.~Bonciani, T.~Gehrmann, R.~Heinesch, P.~Mastrolia and E.~Remiddi,
  %``Decays of scalar and pseudoscalar Higgs bosons into fermions: Two-loop  QCD
  %corrections to the Higgs-quark-antiquark amplitude,''
  Phys.\ Rev.\  D {\bf 72} (2005) 096002
  [arXiv:hep-ph/0508254];
  %%CITATION = PHRVA,D72,096002;%%
%
%\cite{Bernreuther:2004th}
%\bibitem{Bernreuther:2004th}
%  W.~Bernreuther, R.~Bonciani, T.~Gehrmann, R.~Heinesch, T.~Leineweber, P.~Mastrolia and E.~Remiddi,
  %``Two-loop QCD corrections to the heavy quark form factors: Axial vector
  %contributions,''
  Nucl.\ Phys.\  B {\bf 712} (2005) 229
  [arXiv:hep-ph/0412259];
  %%CITATION = NUPHA,B712,229;%%
%
%\cite{Bernreuther:2004ih}
%\bibitem{Bernreuther:2004ih}
%  W.~Bernreuther, R.~Bonciani, T.~Gehrmann, R.~Heinesch, T.~Leineweber, P.~Mastrolia and E.~Remiddi,
  %``Two-loop QCD corrections to the heavy quark form factors: The vector
  %contributions,''
  Nucl.\ Phys.\  B {\bf 706} (2005) 245
  [arXiv:hep-ph/0406046].
  %%CITATION = NUPHA,B706,245;%%

%\cite{Bernreuther:2005rw}
\bibitem{Bernreuther:2005rw}
  W.~Bernreuther, R.~Bonciani, T.~Gehrmann, R.~Heinesch, T.~Leineweber and
E.~Remiddi,
  %``Two-loop QCD corrections to the heavy quark form factors: Anomaly
  %contributions,''
  Nucl.\ Phys.\  B {\bf 723} (2005) 91
  [arXiv:hep-ph/0504190].
  %%CITATION = NUPHA,B723,91;%%

%\cite{Moch:2005tm}
\bibitem{Moch:2005tm}
  S.~Moch, J.~A.~M.~Vermaseren and A.~Vogt,
  %``Three-loop results for quark and gluon form factors,''
  Phys.\ Lett.\  B {\bf 625} (2005) 245
  [arXiv:hep-ph/0508055];
  %%CITATION = PHLTA,B625,245;%%
%
%\cite{Moch:2005id}
%\bibitem{Moch:2005id}
%  S.~Moch, J.~A.~M.~Vermaseren and A.~Vogt,
  %``The quark form factor at higher orders,''
  JHEP {\bf 0508} (2005) 049
  [arXiv:hep-ph/0507039].
  %%CITATION = JHEPA,0508,049;%%

%\cite{Gonsalves:1985jn}
\bibitem{Gonsalves:1985jn}
  R.~J.~Gonsalves,
  %``QCD Electroweak Interference In Vector Boson Production,''
  Phys.\ Rev.\ Lett.\  {\bf 56} (1986) 1647.
  %%CITATION = PRLTA,56,1647;%%

%\cite{Gonsalves:1993sz}
\bibitem{Gonsalves:1993sz}
  R.~J.~Gonsalves and C.~F.~Wai,
  %``Chromoelectroweak Interference And Parity Violating Asymmetries In The
  %Production Of An Electroweak Boson + Two Jets In Hadron Collisions,''
  Phys.\ Rev.\  D {\bf 49} (1994) 190;
  Erratum, {\it ibid}.\  D {\bf 51} (1995) 1428.
  [arXiv:hep-ph/9610278].
  %%CITATION = PHRVA,D49,190;%%

%\cite{Raina:2002uc}
\bibitem{Raina:2002uc}
  S.~Raina,
  ``Interference phenomena in electron positron annihilation,''
  Ph.D.\ Thesis, University at Buffalo, 2002, UMI-30-52534.
  %%CITATION = UMI-30-52534;%%

%\cite{Ellis:1981hk}
\bibitem{Ellis:1981hk}
  R.~K.~Ellis, G.~Martinelli and R.~Petronzio,
  %``Lepton Pair Production At Large Transverse Momentum In Second Order QCD,''
  Nucl.\ Phys.\  B {\bf 211} (1983) 106.
  %%CITATION = NUPHA,B211,106;%%

%\cite{Gonsalves:1989ar}
\bibitem{Gonsalves:1989ar}
  R.~J.~Gonsalves, J.~Pawlowski and C.~F.~Wai,
  %``QCD RADIATIVE CORRECTIONS TO ELECTROWEAK BOSON PRODUCTION AT LARGE
  %TRANSVERSE MOMENTUM IN HADRON COLLISIONS,''
  Phys.\ Rev.\  D {\bf 40} (1989) 2245.
  %%CITATION = PHRVA,D40,2245;%%

%\cite{Arnold:1988dp}
\bibitem{Arnold:1988dp}
  P.~B.~Arnold and M.~H.~Reno,
  %``The Complete Computation of High p(t) W and Z Production in 2nd Order
  %QCD,''
  Nucl.\ Phys.\  B {\bf 319} (1989) 37;
  Erratum, {\it ibid}.\  B {\bf 330} (1990) 284.
  %%CITATION = NUPHA,B319,37;%%

%\cite{Gonsalves:1990ae}
\bibitem{Gonsalves:1990ae}
  R.~J.~Gonsalves, J.~Pawlowski and C.~F.~Wai,
  %``Scale Dependence Of The W And Z Distributions At Large Transverse
  %Momentum,''
  Phys.\ Lett.\  B {\bf 252} (1990) 663.
  %%CITATION = PHLTA,B252,663;%%

%\cite{Albajar:1987aj}
\bibitem{Albajar:1987aj}
  C.~Albajar {\it et al.}  [UA1 Collaboration],
  %``PRODUCTION OF W's WITH LARGE TRANSVERSE MOMENTUM AT THE CERN PROTON -
  %ANTI-PROTON COLLIDER,''
  Phys.\ Lett.\  B {\bf 193} (1987) 389.
  %%CITATION = PHLTA,B193,389;%%

%\cite{Ansari:1988sv}
\bibitem{Ansari:1988sv}
  R.~Ansari {\it et al.}  [UA2 Collaboration],
  %``MEASUREMENT OF THE STRONG COUPLING CONSTANT alpha-s FROM A STUDY OF W
  %BOSONS PRODUCED IN ASSOCIATION WITH JETS,''
  Phys.\ Lett.\  B {\bf 215} (1988) 175.
  %%CITATION = PHLTA,B215,175;%%

%\cite{Martin:1988aj}
\bibitem{Martin:1988aj}
  A.~D.~Martin, R.~G.~Roberts and W.~J.~Stirling,
  %``Improved Parton Distributions And W, Z Production At P Anti-P Colliders,''
  Mod.\ Phys.\ Lett.\  A {\bf 4} (1989) 1135.
  %%CITATION = MPLAE,A4,1135;%%

%\cite{Abbott:2000xv}
\bibitem{Abbott:2000xv}
  B.~Abbott {\it et al.}  [D0 Collaboration],
  %``Differential cross section for $W$ boson production as a function of
  %transverse momentum in $p\bar{p}$ collisions at $\sqrt{s} = 1.8$ TeV,''
  Phys.\ Lett.\  B {\bf 513} (2001) 292
  [arXiv:hep-ex/0010026].
  %%CITATION = PHLTA,B513,292;%%

%\cite{Gonsalves:2005ng}
\bibitem{Gonsalves:2005ng}
  R.~J.~Gonsalves, N.~Kidonakis and A.~S.~Vera,
  %``W production at large transverse momentum at the Large Hadron Collider,''
  Phys.\ Rev.\ Lett.\  {\bf 95} (2005) 222001
  [arXiv:hep-ph/0507317].
  %%CITATION = PRLTA,95,222001;%%

%\cite{GehrmannDeRidder:2008ug}
\bibitem{GehrmannDeRidder:2008ug}
  A.~Gehrmann-De Ridder, T.~Gehrmann, E.~W.~N.~Glover and G.~Heinrich,
  {\it Jet rates in electron-positron annihilation at $O(\alpha_s^3)$ in QCD},
  arXiv:0802.0813 [hep-ph].
  %%CITATION = ARXIV:0802.0813;%%

%\cite{Adler:1969gk}
\bibitem{Adler:1969gk}
  S.~L.~Adler,
  %``Axial vector vertex in spinor electrodynamics,''
  Phys.\ Rev.\  {\bf 177} (1969) 2426.
  %%CITATION = PHRVA,177,2426;%%

%\cite{Bell:1969ts}
\bibitem{Bell:1969ts}
  J.~S.~Bell and R.~Jackiw,
  %``A PCAC puzzle: pi0 $\to$ gamma gamma in the sigma model,''
  Nuovo Cim.\  A {\bf 60} (1969) 47.
  %%CITATION = NUCIA,A60,47;%%

%\cite{Ellis:1985jg}
\bibitem{Ellis:1985jg}
  R.~K.~Ellis and R.~J.~Gonsalves, in
  %``Compact Covariant Cross Sections For Vector Boson Production,''
  {\it Proceedings of the Oregon Workshop on Super High Energy Physics},
  ed. D.E.~Soper (World Scientific, Singapore, 1986), pp. 287-293.
  %%CITATION = C85/08/09;%%

%\cite{Dicus:1985wx}
\bibitem{Dicus:1985wx}
  D.~A.~Dicus and S.~S.~D.~Willenbrock,
  %``Radiative Corrections To The Ratio Of Z And W Boson Production,''
  Phys.\ Rev.\  D {\bf 34} (1986) 148.
  %%CITATION = PHRVA,D34,148;%%

%\cite{Gonsalves:1991qn}
\bibitem{Gonsalves:1991qn}
  R.~J.~Gonsalves, C.~M.~Hung and J.~Pawlowski,
  %``Heavy quark triangle diagram contributions to Z boson production in hadron
  %collisions,''
  Phys.\ Rev.\  D {\bf 46} (1992) 4930.
  %%CITATION = PHRVA,D46,4930;%%

%\cite{Rijken:1995gi}
\bibitem{Rijken:1995gi}
  P.~J.~Rijken and W.~L.~van Neerven,
  %``Heavy Flavor Contributions To The Drell-Yan Cross-Section,''
  Phys.\ Rev.\  D {\bf 52} (1995) 149
  [arXiv:hep-ph/9501373].
  %%CITATION = PHRVA,D52,149;%%

%\cite{Martin:2002aw}
\bibitem{Martin:2002aw}
  A.~D.~Martin, R.~G.~Roberts, W.~J.~Stirling and R.~S.~Thorne,
  %``Uncertainties of predictions from parton distributions. I: Experimental
  %errors. ((T)),''
  Eur.\ Phys.\ J.\  C {\bf 28} (2003) 455
  [arXiv:hep-ph/0211080].
  %%CITATION = EPHJA,C28,455;%%

\end{thebibliography}
\end{document}